\documentclass[twocolumn,superscriptaddress,showpacs,preprintnumbers,amsmath,amssymb,aps,prb]{revtex4}
\usepackage{graphicx}% Include figure files
\usepackage{dcolumn}% Align table columns on decimal point
\usepackage{bm}% bold math
\begin{document}
\title{Interconnection of point defect parameters in solids with bulk properties: Application to diamond}
\author{P. A. Varotsos}
\email{pvaro@otenet.gr} \affiliation{Solid State Section, Physics
Department, University of Athens, Panepistimiopolis, Zografos 157
84, Athens, Greece} \affiliation{Solid Earth Physics Institute,
Physics Department, University of Athens, Panepistimiopolis,
Zografos 157 84, Athens, Greece}

\begin{abstract}
 We show that the  values of the defect entropy and the
 defect enthalpy for the vacancy formation in diamond, have a
 ratio which is comparable to the one predicted by a model
 suggested four decades ago. This
 model, which interconnects the formation Gibbs energy with the
 bulk elastic and expansivity data, has been also recently found of
 value in high $T_c$-superconductors as well as in glass-forming
 liquids.
\end{abstract}
\maketitle

The following interconnection between the Gibbs energy $g^f$ for
the defect formation ($f$), or the Gibbs energy $g^m$ for the
defect migration ($m$), and the bulk properties (i.e., the
expansivity and the elastic data) has been suggested long
ago\cite{VARALEX80,VARALEX81,varbook}:
\begin{equation}\label{eq1}
    g^f=c^fB\Omega
\end{equation}
and
\begin{equation}\label{eq2}
    g^m=c^mB\Omega
\end{equation}
(called $cB\Omega$ model) where $B$ is the isothermal bulk modulus
and $\Omega$ the mean volume per atom. It has been
argued\cite{varbook} that the values of $c^f$ and $c^m$ are
practically {\em independent} of temperature and pressure, As a
consequence, the corresponding Gibbs energy $g^{act}$ for
activation (act) processes (when a single mechanism is operating, e.g. for conductivity \cite{VAR74,KOS75})
is given by
\begin{equation}\label{eq3}
    g^{act}=c^{act}B\Omega
\end{equation}
where $c^{act}(=c^f+c^m)$ is also independent of temperature and
pressure. (cf. As an example, Eq.(\ref{eq3}) enables\cite{varbook}
the calculation of self-diffusion coefficients at any temperature
and pressure from a single measurement.) For reasons of brevity,
the above relations are jointly written as
\begin{equation}\label{eq4}
    g^i=c^iB\Omega
\end{equation}
where $i$ stands for the corresponding process, i.e., $i=f,m,act$.
By differentiating Eq.(\ref{eq4}) in respect to either the
temperature $T$ for ($P$=const) or the pressure $P$ for
($T$=const), we find the following expressions for the defect
entropy $s^i$ ($=-  \left. \frac{dg^i}{dT} \right|_P $), the defect
enthalpy $h^i$($=g^i-T\left. \frac{dg^i}{dT} \right|_P$) and the
defect volume $v^i$ ($=\left. \frac{dg^i}{dP} \right|_T$):

\begin{equation}\label{eq5}
s^i=-c^i \Omega \left( \beta B + \left. \frac{dB}{dT} \right|_P
\right)
\end{equation}

\begin{equation}\label{eq6}
h^i=c^i \Omega \left( B -T \beta B -T \left. \frac{dB}{dT}
\right|_P \right)
\end{equation}

\begin{equation}\label{eq7}
v^i=c^i \Omega \left( \left. \frac{dB}{dP} \right|_T -1 \right)
\end{equation}
where $\beta$ stands for the thermal (volume) expansion
coefficient. These relations reveal that the ratios $s^i/h^i$ and
$v^i/h^i$ are {\em solely} governed by the macroscopic properties
$\beta$, $B$, $\left. \frac{dB}{dT} \right|_P$ and $\left.
\frac{dB}{dP} \right|_T$. This has been checked  in a variety of
solids, i.e., rare gas
solids, metals and ionic crystals, and the results have been
compiled in a monograph.\cite{varbook} Furthermore, the $cB\Omega$
model enables\cite{varbook} the determination of the concentration
at which the conductivity $(\sigma)$ and the self-diffusion
coefficients ($D$) maximize when studying mixed alkali halides
(solid solutions) (cf. In the latter solids, the values of the
dielectric constant $\epsilon$ of the mixed crystal (if they are not experimentally available) can be
estimated\cite{VAR80K133} at various concentrations in terms of the
$\epsilon$ values of the pure constituents).

Later\cite{var99} the $cB\Omega$ model was found of value to
describe experimental results in ionic crystals, which revealed
that a time dependent electric polarization arises (in the {\em absence}
of any external electric field) upon changing the rate of the
uniaxial stress or by the indenter penetration into the crystal
surface. These experiments led to activation volumes which
-although being order(s) of magnitude smaller than those measured
in other cases, e.g., in alkali halides- were found to
agree\cite{var99} with the ratio $v^i/h^i$ predicted from the
$cB\Omega$ model. Such experiments have a specific importance,
because they are closely interrelated \cite{var99} with transient
electric signals that have been found to
precede\cite{EPL12,VAR96DEB,JAP2} earthquakes (cf. these signals are
emitted from the focal region, where a stress accumulation is
expected\cite{varbook} to occur before the earthquake occurrence).

The interconnection between $g^i$ and elastic data has recently
attracted a strong interest, in view of a number of challenging
findings, chief among of which are the following:

First, in type-II superconductors, high critical current densities
can be achieved by the presence of high-density defects which will
provide suitable pinning centers for the magnetic flux lines
(e.g., see Ref.[\onlinecite{su04}] and reference therein). In such
a flux-line pinning, point defects can play an important role in
cuprates such as YBa$_2$Cu$_3$O$_{7-\delta}$ (Y-123) in which the
superconducting coherence length is of the order of tens of \AA.
Doping YBa$_2$Cu$_3$O$_{7-\delta}$ with alkaline earth elements
improve bulk and grain boundary transport and other properties
(e.g., Ref.[\onlinecite{hil02}] and references therein).
Interestingly, it was found\cite{su04} that in the latter
superconductors, the formation volume of Schottky defects obeys the behavior predicted by
the $cB\Omega$-model. On the other hand, the model does not seem
to describe the case when the replacement of $Y$ is made by
rare-earth elements. (cf. This is strikingly reminiscent of the
early finding\cite{var74aaa} that in alkali halides doped with
alkaline earth divalent cations, the $h^m$-value for the
reorientation process of the electric dipoles formed by  a
divalent cation and a nearby cation vacancy  (taking place
through jumps of the cation vacancy neighboring of the divalent impurity), increases upon
increasing the ionic radius of the dopants; this is {\em not} the
case, however, for rare-earth divalent dopants. The origin of this
striking analogy
has not yet been understood).

Second, when studying the basic characteristics of the
liquid-glass transitions, almost  all glass-forming liquids
(including oxide melts, ionic liquids, metallic liquid alloys,
polymers, molecular liquids or viscous liquids studied by computer
simulations) exhibit ``universal'' features which are not yet well
understood.\cite{dyr06a} Chief among these features is the fact
that the viscosity of glass-forming liquids with few exceptions
exhibits a temperature dependence which markedly deviates from the
Arrhenius behavior. This has been recently investigated in depth
in a series of challenging papers\cite{dyr06a,dyr06b,dyr06c} which
point to the conclusion that the elastic models play a prominent
role in the following sense. Following Ref.[\onlinecite{dyr06c}],
let us start from the point that viscous liquids could be viewed
more as ``solids which flow'' than as less-viscous liquids like
ambient water; the glass-forming liquids exhibit extremely large
viscosity upon approaching the glass transition, thus most
molecular motion goes into vibrations, just like in a solid. Only
rarely does anything happen in the form of a flow event, a
molecular rearrangement (in which the migration barrier is $\gg
k_BT$, where $k_B$ stands for the usual Boltzmann constant). These
flow events are similar to the point defect motion in solids and,
in the frame of the elastic models, the non-Arrhenius behavior of
the viscosity is captured by the temperature variation of the
elastic constants.

The present study is focused on the investigation of the validity
of the $cB\Omega$-model in  diamond, which exhibits certain
properties that differ significantly from those in the materials
mentioned above, e.g., it has a very large Debye temperature,
$\Theta_D \approx 2246K$, making it a ``quantum'' crystal even at
room temperature. Its hardness and abrasive qualities, highly
valued in technology and gem industry, are controlled by the large
elastic moduli (e.g., see Ref.[\onlinecite{zou98}] and references
therein). In general, the point defect parameters of diamond are
of great interest in diverse fields. For example, in earth
sciences its diffusion properties have a specific importance,
because natural diamonds and their mineral inclusions provide
information about the geochemical character and geotherm of the
ancient continental lithosphere (e.g., see
Ref.[\onlinecite{kog05}] and references therein). In particular,
the spatial distribution of carbon and nitrogen isotopes in
diamond provides information on mantle residence time\cite{kog03}.
Diamonds with long residence at high temperature will gradually
lose their initial zoning patterns due to diffusion and hence the
diffusion data can constrain the maximum possible age of diamonds.
Rare diamonds originating from the mantle transition zone should
have developed\cite{kog03} a length scale of $\approx$1mm of
isotopic zoning over the age of the Earth ($4.5\times10^9$ years).

A large body of data has been accumulated in diamond during the
last decade, which allowed the present study to become
possible. First, the $B$-values have been calculated by a general
model\cite{agu06} (relating mechanical and vibrational properties
through a combination of first Szigeti and Lyddane-Sachs-Teller
relations) up to 1800$K$ from available Raman
measurements\cite{her91,zou91,cui98,liu00}. The calculated values
are in excellent agreement with experimental results \cite{zou98}
up to 1600$K$ that were based on Brillouin scattering
measurements. Second, the vacancy formation enthalpy was
calculated to be $h^f=7.2eV$ by Brenner et al.\cite{bre02} (see
their Table 10). Third, by using the self-consistent charge
density-functional based tight- binding method and correcting for
the strong finite-size effects, Rauls and Frauenheim\cite{rau04}
achieved in finding that the vacancy formation entropy is
$s^f=2.85k_B$ with a plausible uncertainty which is less than
$\pm0.3k_B$. Finally, the self-diffusion coefficients have been
  measured\cite{kog05} in a natural diamond at the
condition within its field of stability: 10GPa and 2075-2375$K$.
An activation enthalpy $h^{act}=6.8\pm1.6eV$ was then reported,
which is approximately 30\% lower than results predicted from {\em
ab initio} calculations\cite{ber88,HOO3}.

We now proceed to a numerical check of the $cB\Omega$ model. A
combination of Eqs.(\ref{eq5}) and (\ref{eq6}), for $i=f$, reveals:\cite{varbook}
\begin{equation}\label{eq8}
\frac{s^f}{h^f}= - \frac{\beta B+\left. \frac{dB}{dT}
\right|_P}{B-T \beta B - T \left. \frac{dB}{dT} \right|_P}
\end{equation}
Let us make the calculation for the highest temperature $T$=1600K
at which, according to Aguado and Baonza,\cite{agu06} we can rely on the
experimental $B$-values. A least squares fit to a
straight line of the $B$-values that are given in Fig.1 of
Ref.[\onlinecite{agu06}] as a function of $T$, gives, in the range
 $T \geq1200K$,
the value $\left. \frac{dB}{dT} \right|_P \approx -2.5 \times
10^{-2}$ MPa/K (with a plausible uncertainty of around 10\%), while the
$B$-value (at 1600K) is $\approx 415$ GPa. Taking the density from the
expression\cite{zou98} $\rho =3.513+7.4 \times 10^{-6}T-3.8 \times
10^{-8}T^2+7.1 \times 10^{-12}T^3$ as deduced from the thermal
expansion data of Slack and Bartman\cite{sla75}, we estimate
(for $T=1600K$) that $\beta \approx 17.26 \times 10^{-6}K^{-1}$. By
inserting these values into Eq.(\ref{eq8}) we find
\begin{equation}\label{eq9}
    \frac{s^f}{h^f}\approx 4\times10^{-5} K^{-1}
\end{equation}
(with a plausible uncertainty of around 10\%). We now compare this result  with
the value
\begin{equation}\label{eq10}
 \frac{s^f}{h^f}=3.4^{+0.4}_{-0.4} \times 10^{-5} K^{-1}
\end{equation}
deduced when inserting the published parameters
$s^f=(2.85\pm0.3)k_B$ (from Ref.[\onlinecite{rau04}]) and
$h^f=7.2eV$ (from Ref.[\onlinecite{bre02}]) mentioned above. This
comparison indicates more or less a satisfactory agreement, if one
also considers the uncertainties involved. The agreement becomes
even better if one takes into account that, in the light of the
aforementioned self-diffusion experimental results which led to
$h^{act}(=h^m+h^f)\approx 6.8eV$, the value of $7.2eV$ used in
Eq.(\ref{eq10}) seems to somewhat overestimate the actual
$h^f$-value. Furthermore, we note that the experimental value
$h^{act}\approx 6.8eV$ enables  the direct determination of
$v^{act}$ (for the self diffusion process) as follows: By dividing
Eqs.(\ref{eq6}) and (\ref{eq7}), we immediately find the ratio
$v^{act}/h^{act}$, i.e.,
\begin{equation}\label{eq9}
\frac{v^f}{h^f}= \frac{\frac{dB}{dP}|_T-1} {B-T \beta B - T \left.
\frac{dB}{dT} \right|_P}
\end{equation}
Upon inserting the aforementioned elastic and expansivity data
along with the value\cite{gil99} $\left. \frac{dB}{dP}
\right|_T\approx 4$ and then using the value $h^{act}\approx
6.8eV$, we get $v^{act} \approx 4.4  cm^3mol^{-1}$. Unfortunately,
a comparison of this calculated $v^{act}$ value with an
experimental result cannot be made, because to the best of our
knowledge an experiment towards determining $v^{act}$ (i.e.,
self-diffusion measurements at various pressures) has not yet been
performed.

%\bibliographystyle{apsrev}
%\bibliography{pvaroprb}

\begin{thebibliography}{30}
\expandafter\ifx\csname natexlab\endcsname\relax\def\natexlab#1{#1}\fi
\expandafter\ifx\csname bibnamefont\endcsname\relax
  \def\bibnamefont#1{#1}\fi
\expandafter\ifx\csname bibfnamefont\endcsname\relax
  \def\bibfnamefont#1{#1}\fi
\expandafter\ifx\csname citenamefont\endcsname\relax
  \def\citenamefont#1{#1}\fi
\expandafter\ifx\csname url\endcsname\relax
  \def\url#1{\texttt{#1}}\fi
\expandafter\ifx\csname urlprefix\endcsname\relax\def\urlprefix{URL }\fi
\providecommand{\bibinfo}[2]{#2}
\providecommand{\eprint}[2][]{\url{#2}}

\bibitem[{\citenamefont{Varotsos and Alexopoulos}(1980)}]{VARALEX80}
\bibinfo{author}{\bibfnamefont{P.}~\bibnamefont{Varotsos}} \bibnamefont{and}
  \bibinfo{author}{\bibfnamefont{K.}~\bibnamefont{Alexopoulos}},
  \bibinfo{journal}{J. Phys. Chem. Sol.} \textbf{\bibinfo{volume}{41}},
  \bibinfo{pages}{443} (\bibinfo{year}{1980}).

\bibitem[{\citenamefont{Varotsos and Alexopoulos}(1981)}]{VARALEX81}
\bibinfo{author}{\bibfnamefont{P.}~\bibnamefont{Varotsos}} \bibnamefont{and}
  \bibinfo{author}{\bibfnamefont{K.}~\bibnamefont{Alexopoulos}},
  \bibinfo{journal}{J. Phys. Chem. Sol.} \textbf{\bibinfo{volume}{42}},
  \bibinfo{pages}{409} (\bibinfo{year}{1981}).

\bibitem[{\citenamefont{Varotsos and Alexopoulos}(1986)}]{varbook}
\bibinfo{author}{\bibfnamefont{P.}~\bibnamefont{Varotsos}} \bibnamefont{and}
  \bibinfo{author}{\bibfnamefont{K.}~\bibnamefont{Alexopoulos}},
  \emph{\bibinfo{title}{Thermodynamics of Point Defects and their Relation with
  Bulk Properties}} (\bibinfo{publisher}{North Holland},
  \bibinfo{address}{Amsterdam}, \bibinfo{year}{1986}).

\bibitem[{\citenamefont{Varotsos and Mourikis}(1974)}]{VAR74}
\bibinfo{author}{\bibfnamefont{P.}~\bibnamefont{Varotsos}} \bibnamefont{and}
  \bibinfo{author}{\bibfnamefont{S.}~\bibnamefont{Mourikis}},
  \bibinfo{journal}{Phys. Rev. B} \textbf{\bibinfo{volume}{10}},
  \bibinfo{pages}{5220} (\bibinfo{year}{1974}).

\bibitem[{\citenamefont{Kostopoulos et~al.}(1975)\citenamefont{Kostopoulos,
  Varotsos, and Mourikis}}]{KOS75}
\bibinfo{author}{\bibfnamefont{D.}~\bibnamefont{Kostopoulos}},
  \bibinfo{author}{\bibfnamefont{P.}~\bibnamefont{Varotsos}}, \bibnamefont{and}
  \bibinfo{author}{\bibfnamefont{S.}~\bibnamefont{Mourikis}},
  \bibinfo{journal}{Can. J. Phys.} \textbf{\bibinfo{volume}{53}},
  \bibinfo{pages}{1318 } (\bibinfo{year}{1975}).

\bibitem[{\citenamefont{Varotsos}(1980)}]{VAR80K133}
\bibinfo{author}{\bibfnamefont{P.}~\bibnamefont{Varotsos}},
  \bibinfo{journal}{physica status solidi (b)} \textbf{\bibinfo{volume}{100}},
  \bibinfo{pages}{K133} (\bibinfo{year}{1980}).

\bibitem[{\citenamefont{Varotsos et~al.}(1999)\citenamefont{Varotsos, Sarlis,
  and Lazaridou}}]{var99}
\bibinfo{author}{\bibfnamefont{P.}~\bibnamefont{Varotsos}},
  \bibinfo{author}{\bibfnamefont{N.}~\bibnamefont{Sarlis}}, \bibnamefont{and}
  \bibinfo{author}{\bibfnamefont{M.}~\bibnamefont{Lazaridou}},
  \bibinfo{journal}{Phys. Rev. B} \textbf{\bibinfo{volume}{59}},
  \bibinfo{pages}{24} (\bibinfo{year}{1999}).

\bibitem[{\citenamefont{Varotsos et~al.}(2012)\citenamefont{Varotsos, Sarlis,
  and Skordas}}]{EPL12}
\bibinfo{author}{\bibfnamefont{P.}~\bibnamefont{Varotsos}},
  \bibinfo{author}{\bibfnamefont{N.}~\bibnamefont{Sarlis}}, \bibnamefont{and}
  \bibinfo{author}{\bibfnamefont{E.}~\bibnamefont{Skordas}},
  \bibinfo{journal}{EPL (Europhysics Letters)} \textbf{\bibinfo{volume}{99}}, \bibinfo{eid}{59001}
  (\bibinfo{year}{2012}).

\bibitem[{\citenamefont{Varotsos et~al.}(1996)\citenamefont{Varotsos, Eftaxias,
  Vallianatos, and Lazaridou}}]{VAR96DEB}
\bibinfo{author}{\bibfnamefont{P.}~\bibnamefont{Varotsos}},
  \bibinfo{author}{\bibfnamefont{K.}~\bibnamefont{Eftaxias}},
  \bibinfo{author}{\bibfnamefont{F.}~\bibnamefont{Vallianatos}},
  \bibnamefont{and}
  \bibinfo{author}{\bibfnamefont{M.}~\bibnamefont{Lazaridou}},
  \bibinfo{journal}{Geophys. Res. Lett.} \textbf{\bibinfo{volume}{23}},
  \bibinfo{pages}{1295} (\bibinfo{year}{1996}).

\bibitem[{\citenamefont{Varotsos et~al.}(2000)\citenamefont{Varotsos, Sarlis,
  and Lazaridou}}]{JAP2}
\bibinfo{author}{\bibfnamefont{P.}~\bibnamefont{Varotsos}},
  \bibinfo{author}{\bibfnamefont{N.}~\bibnamefont{Sarlis}}, \bibnamefont{and}
  \bibinfo{author}{\bibfnamefont{M.}~\bibnamefont{Lazaridou}},
  \bibinfo{journal}{Acta Geophysica Polonica} \textbf{\bibinfo{volume}{48}},
  \bibinfo{pages}{141} (\bibinfo{year}{2000}).

\bibitem[{\citenamefont{Su et~al.}(2004)\citenamefont{Su, Welch, and
  Wong-Ng}}]{su04}
\bibinfo{author}{\bibfnamefont{H.}~\bibnamefont{Su}},
  \bibinfo{author}{\bibfnamefont{D.~O.} \bibnamefont{Welch}}, \bibnamefont{and}
  \bibinfo{author}{\bibfnamefont{W.}~\bibnamefont{Wong-Ng}},
  \bibinfo{journal}{Phys. Rev. B} \textbf{\bibinfo{volume}{70}},
  \bibinfo{pages}{054517} (\bibinfo{year}{2004}).

\bibitem[{\citenamefont{Hilgenkamp and Mannhart}(2002)}]{hil02}
\bibinfo{author}{\bibfnamefont{H.}~\bibnamefont{Hilgenkamp}} \bibnamefont{and}
  \bibinfo{author}{\bibfnamefont{J.}~\bibnamefont{Mannhart}},
  \bibinfo{journal}{Rev. Mod. Phys.} \textbf{\bibinfo{volume}{74}},
  \bibinfo{pages}{485} (\bibinfo{year}{2002}).

\bibitem[{\citenamefont{Varotsos and Miliotis}(1974)}]{var74aaa}
\bibinfo{author}{\bibfnamefont{P.}~\bibnamefont{Varotsos}} \bibnamefont{and}
  \bibinfo{author}{\bibfnamefont{D.}~\bibnamefont{Miliotis}},
  \bibinfo{journal}{J. Phys. Chem. Sol.} \textbf{\bibinfo{volume}{35}},
  \bibinfo{pages}{927} (\bibinfo{year}{1974}).

\bibitem[{\citenamefont{Dyre}(2006{\natexlab{a}})}]{dyr06a}
\bibinfo{author}{\bibfnamefont{J.~C.} \bibnamefont{Dyre}},
  \bibinfo{journal}{Rev. Mod. Phys.} \textbf{\bibinfo{volume}{78}},
  \bibinfo{pages}{953} (\bibinfo{year}{2006}{\natexlab{a}}).

\bibitem[{\citenamefont{Dyre}(2006{\natexlab{b}})}]{dyr06b}
\bibinfo{author}{\bibfnamefont{J.~C.} \bibnamefont{Dyre}},
  \bibinfo{journal}{AIP Conference Proceedings} \textbf{\bibinfo{volume}{832}},
  \bibinfo{pages}{113} (\bibinfo{year}{2006}{\natexlab{b}}).

\bibitem[{\citenamefont{Dyre et~al.}(2006)\citenamefont{Dyre, Christensen, and
  Olsen}}]{dyr06c}
\bibinfo{author}{\bibfnamefont{J.~C.} \bibnamefont{Dyre}},
  \bibinfo{author}{\bibfnamefont{T.}~\bibnamefont{Christensen}},
  \bibnamefont{and} \bibinfo{author}{\bibfnamefont{N.~B.} \bibnamefont{Olsen}},
  \bibinfo{journal}{J. Non-Cryst. Solids} \textbf{\bibinfo{volume}{352}},
  \bibinfo{pages}{4635} (\bibinfo{year}{2006}).

\bibitem[{\citenamefont{Zouboulis et~al.}(1998)\citenamefont{Zouboulis,
  Grimsditch, Ramdas, and Rodriguez}}]{zou98}
\bibinfo{author}{\bibfnamefont{E.~S.} \bibnamefont{Zouboulis}},
  \bibinfo{author}{\bibfnamefont{M.}~\bibnamefont{Grimsditch}},
  \bibinfo{author}{\bibfnamefont{A.~K.} \bibnamefont{Ramdas}},
  \bibnamefont{and}
  \bibinfo{author}{\bibfnamefont{S.}~\bibnamefont{Rodriguez}},
  \bibinfo{journal}{Phys. Rev. B} \textbf{\bibinfo{volume}{57}},
  \bibinfo{pages}{2889} (\bibinfo{year}{1998}).

\bibitem[{\citenamefont{Koga et~al.}(2005)\citenamefont{Koga, Walter, Nakumura,
  and Kobayashi}}]{kog05}
\bibinfo{author}{\bibfnamefont{K.~T.} \bibnamefont{Koga}},
  \bibinfo{author}{\bibfnamefont{M.~J.} \bibnamefont{Walter}},
  \bibinfo{author}{\bibfnamefont{E.}~\bibnamefont{Nakumura}}, \bibnamefont{and}
  \bibinfo{author}{\bibfnamefont{K.}~\bibnamefont{Kobayashi}},
  \bibinfo{journal}{Phys. Rev. B} \textbf{\bibinfo{volume}{72}},
  \bibinfo{pages}{024108} (\bibinfo{year}{2005}).

\bibitem[{\citenamefont{Koga et~al.}(2003)\citenamefont{Koga, Van~Orman, and
  Walter}}]{kog03}
\bibinfo{author}{\bibfnamefont{K.~T.} \bibnamefont{Koga}},
  \bibinfo{author}{\bibfnamefont{J.~A.} \bibnamefont{Van~Orman}},
  \bibnamefont{and} \bibinfo{author}{\bibfnamefont{M.~J.}
  \bibnamefont{Walter}}, \bibinfo{journal}{Phys. Earth Planet. Int.}
  \textbf{\bibinfo{volume}{139}}, \bibinfo{pages}{35} (\bibinfo{year}{2003}).

\bibitem[{\citenamefont{Aguado and Baonza}(2006)}]{agu06}
\bibinfo{author}{\bibfnamefont{F.}~\bibnamefont{Aguado}} \bibnamefont{and}
  \bibinfo{author}{\bibfnamefont{V.~G.} \bibnamefont{Baonza}},
  \bibinfo{journal}{Phys. Rev. B} \textbf{\bibinfo{volume}{73}},
  \bibinfo{pages}{024111} (\bibinfo{year}{2006}).

\bibitem[{\citenamefont{Herchen and Cappelli}(1991)}]{her91}
\bibinfo{author}{\bibfnamefont{H.}~\bibnamefont{Herchen}} \bibnamefont{and}
  \bibinfo{author}{\bibfnamefont{M.~A.} \bibnamefont{Cappelli}},
  \bibinfo{journal}{Phys. Rev. B} \textbf{\bibinfo{volume}{43}},
  \bibinfo{pages}{11740} (\bibinfo{year}{1991}).

\bibitem[{\citenamefont{Zouboulis and Grimsditch}(1991)}]{zou91}
\bibinfo{author}{\bibfnamefont{E.~S.} \bibnamefont{Zouboulis}}
  \bibnamefont{and}
  \bibinfo{author}{\bibfnamefont{M.}~\bibnamefont{Grimsditch}},
  \bibinfo{journal}{Phys. Rev. B} \textbf{\bibinfo{volume}{43}},
  \bibinfo{pages}{12490} (\bibinfo{year}{1991}).

\bibitem[{\citenamefont{Cui et~al.}(1998)\citenamefont{Cui, Amtmann, Ristein,
  and Ley}}]{cui98}
\bibinfo{author}{\bibfnamefont{J.~B.} \bibnamefont{Cui}},
  \bibinfo{author}{\bibfnamefont{K.}~\bibnamefont{Amtmann}},
  \bibinfo{author}{\bibfnamefont{J.}~\bibnamefont{Ristein}}, \bibnamefont{and}
  \bibinfo{author}{\bibfnamefont{L.}~\bibnamefont{Ley}}, \bibinfo{journal}{J.
  Appl. Phys.} \textbf{\bibinfo{volume}{83}}, \bibinfo{pages}{7929}
  (\bibinfo{year}{1998}).

\bibitem[{\citenamefont{Liu et~al.}(2000)\citenamefont{Liu, Bursill, Prawer,
  and Beserman}}]{liu00}
\bibinfo{author}{\bibfnamefont{M.~S.} \bibnamefont{Liu}},
  \bibinfo{author}{\bibfnamefont{L.~A.} \bibnamefont{Bursill}},
  \bibinfo{author}{\bibfnamefont{S.}~\bibnamefont{Prawer}}, \bibnamefont{and}
  \bibinfo{author}{\bibfnamefont{R.}~\bibnamefont{Beserman}},
  \bibinfo{journal}{Phys. Rev. B} \textbf{\bibinfo{volume}{61}},
  \bibinfo{pages}{3391} (\bibinfo{year}{2000}).

\bibitem[{\citenamefont{Brenner et~al.}(2002)\citenamefont{Brenner, Shenderova,
  Harrison, Stuart, Ni, and Sinnott}}]{bre02}
\bibinfo{author}{\bibfnamefont{D.~W.} \bibnamefont{Brenner}},
  \bibinfo{author}{\bibfnamefont{O.~A.} \bibnamefont{Shenderova}},
  \bibinfo{author}{\bibfnamefont{J.~A.} \bibnamefont{Harrison}},
  \bibinfo{author}{\bibfnamefont{S.~J.} \bibnamefont{Stuart}},
  \bibinfo{author}{\bibfnamefont{B.}~\bibnamefont{Ni}}, \bibnamefont{and}
  \bibinfo{author}{\bibfnamefont{S.~B.} \bibnamefont{Sinnott}},
  \bibinfo{journal}{J. Phys. Condens. Matter} \textbf{\bibinfo{volume}{14}},
  \bibinfo{pages}{783} (\bibinfo{year}{2002}).

\bibitem[{\citenamefont{Rauls and Frauenheim}(2004)}]{rau04}
\bibinfo{author}{\bibfnamefont{E.}~\bibnamefont{Rauls}} \bibnamefont{and}
  \bibinfo{author}{\bibfnamefont{T.}~\bibnamefont{Frauenheim}},
  \bibinfo{journal}{Phys. Rev. B} \textbf{\bibinfo{volume}{69}},
  \bibinfo{pages}{155213} (\bibinfo{year}{2004}).

\bibitem[{\citenamefont{Bernholc et~al.}(1988)\citenamefont{Bernholc,
  Antonelli, Del~Sole, Bar-Yam, and Pantelides}}]{ber88}
\bibinfo{author}{\bibfnamefont{J.}~\bibnamefont{Bernholc}},
  \bibinfo{author}{\bibfnamefont{A.}~\bibnamefont{Antonelli}},
  \bibinfo{author}{\bibfnamefont{T.~M.} \bibnamefont{Del~Sole}},
  \bibinfo{author}{\bibfnamefont{Y.}~\bibnamefont{Bar-Yam}}, \bibnamefont{and}
  \bibinfo{author}{\bibfnamefont{S.~T.} \bibnamefont{Pantelides}},
  \bibinfo{journal}{Phys. Rev. Lett.} \textbf{\bibinfo{volume}{61}},
  \bibinfo{pages}{2689} (\bibinfo{year}{1988}).

\bibitem[{\citenamefont{Hood et~al.}(2003)\citenamefont{Hood, Kent, Needs, and
  Briddon}}]{HOO3}
\bibinfo{author}{\bibfnamefont{R.~Q.} \bibnamefont{Hood}},
  \bibinfo{author}{\bibfnamefont{P.~R.~C.} \bibnamefont{Kent}},
  \bibinfo{author}{\bibfnamefont{R.~J.} \bibnamefont{Needs}}, \bibnamefont{and}
  \bibinfo{author}{\bibfnamefont{P.~R.} \bibnamefont{Briddon}},
  \bibinfo{journal}{Phys. Rev. Lett.} \textbf{\bibinfo{volume}{91}},
  \bibinfo{pages}{076403} (\bibinfo{year}{2003}).

\bibitem[{\citenamefont{Slack and Bartram}(1975)}]{sla75}
\bibinfo{author}{\bibfnamefont{G.~A.} \bibnamefont{Slack}} \bibnamefont{and}
  \bibinfo{author}{\bibfnamefont{S.~F.} \bibnamefont{Bartram}},
  \bibinfo{journal}{J. Appl. Phys.} \textbf{\bibinfo{volume}{46}},
  \bibinfo{pages}{89} (\bibinfo{year}{1975}).

\bibitem[{\citenamefont{Gillet et~al.}(1999)\citenamefont{Gillet, Fiquet,
  Daniel, Reynard, and Hanfland}}]{gil99}
\bibinfo{author}{\bibfnamefont{P.}~\bibnamefont{Gillet}},
  \bibinfo{author}{\bibfnamefont{G.}~\bibnamefont{Fiquet}},
  \bibinfo{author}{\bibfnamefont{I.}~\bibnamefont{Daniel}},
  \bibinfo{author}{\bibfnamefont{B.}~\bibnamefont{Reynard}}, \bibnamefont{and}
  \bibinfo{author}{\bibfnamefont{M.}~\bibnamefont{Hanfland}},
  \bibinfo{journal}{Phys. Rev. B} \textbf{\bibinfo{volume}{60}},
  \bibinfo{pages}{14660} (\bibinfo{year}{1999}).

\end{thebibliography}

\end{document}